\documentclass[12pt]{iopart}

\begin{document}

\title{Realism and the wave function}
\author{A.\ Matzkin }
\address{Department of Physics and Astronomy, University College London, Gower Street, London WC1E 6BT, UK}

\begin{abstract}
Realism -- the idea that the concepts in physical theories refer
to 'things' existing in the real world -- is introduced as a tool
to analyze the status of the wave-function. Although the physical
entities are recognized by the existence of invariant quantities,
examples from classical and quantum physics suggest that not all
the theoretical terms refer to the entities: some terms refer to
properties of the entities, and some terms have only an epistemic
function. In particular, it is argued that the wave-function may
be written in terms of classical non-referring and epistemic
terms. The implications for realist interpretations of quantum
mechanics and on the teaching of quantum physics are examined.
\end{abstract}

\maketitle

\section{Introduction}

For Weinberg, '{\it wave-functions are real (...) because it is useful to
include them in our theories}' \cite{weinberg92}. Popper on the other hand,
although reluctant to argue about words, believed something is real provided
it is '{\em ''kickable'' and able to kick back if kicked}' \cite{popper}.
There is a persistent confusion in the debate on the reality of quantum
systems, on what a quantum object is, and even whether there are such things
as quantum objects at all. This is not new -- the debate between the
''Founding Fathers'' of 20th century physics is plagued with
misunderstandings arising because of alternative or different meanings
conveyed by key words such as realism, determinism, probability, causality,
etc. It may be noted that still today, it is widely believed that the defeat
of Einstein's position has rang the knell of realism (whereas Einstein's
realism is of a very specific kind), or conversely that the de Broglie-Bohm
version of quantum mechanics is the only one compatible with realism
(whereas history of science taken by and large indicates that the more a
theory contains ad-hoc elements unobservable in principle, the less it
refers to the real world).

Concurrently, recent investigations on the students' understanding of
quantum mechanics have all highlighted profound conceptual problems. But
depending on the authors' conceptions on the status of quantum physics
vis-\`{a}-vis reality, opposite conclusions have been drawn, eg from the
need to '{\it reconcile non-visualizable quantum physics with visualizable
classical physics}' \cite{mash99} to prescription of '{\it avoiding
reference to classical physics}' \cite{ireson99}.

Part of the above-mentioned problems are not so much related to
what realism \emph{is} (there is of course no general agreement
among philosophers on this point), but rather to establishing
clear-cut categories helpful in the analysis of quantum-mechanical
reality. Note that the very nature of the physical reality
described by quantum-mechanics is still today a matter of personal
taste or of philosophical prejudice. The {\em ontological} basis
(ie, what the world is made of) underlying quantum mechanics
depends on the meaning given to the theoretical terms: are these
terms real, or are they just {\em epistemic} (ie, purely
knowledge-related) tools? Our aim here is to draw some simple
conceptual distinctions concerning realism. More specifically, we
will analyze the relation between quantum mechanics and physical
reality from the standpoint of reference. Present-day scientific
realism is generally based on one or other version of a theory of
reference, by which theoretical terms are assumed to refer to
entities existing in the real world (Sec. 2). This doesn't entail
however that {\em every} theoretical term refers to an existing
physical entity. We will argue in Secs. 3 and 4 that the
relationship between the theoretical terms and the physical
entities is a relevant starting point to investigate the status of
the theoretical terms. We shall take examples of referring and
non-referring theoretical terms from classical mechanics, and
examine the situation in quantum mechanics; in particular, we will
consider whether the wave-function can be said to be an objective
physical system, as has been recently suggested \cite{paty99}. The
implications of our analysis for realist interpretations of
quantum mechanics and on the students' understanding of quantum
phenomena will be detailed in Sec. 5.

\section{Realism}

As mentioned in the introduction, there is no agreement among the
philosophers of science as to what realism is. In fact, there are very
different sorts of realism, and it is not our aim here to review these (see
eg \cite{putnam89}, and \cite{psillos} and references therein). Here, we
will understand by realism not so much the hypothesis that there is an
autonomous external world known as ''physical reality'' (the overwhelming
majority of physicists would agree with this assumption\footnote{%
There are some well-known exceptions however, such as Wigner \cite{wigner}
or Wheeler \cite{wheeler}.}) but rather the stronger idea that the concepts
of physical theories refer to 'things' existing in the real world, ie
physical theories and concepts are more than convenient manners of
organizing the data obtained from observation. Indeed, if that were to be
the case, that is if physical theories and concepts spoke of objects that do
not really exist, then, as Putnam put it, only a miracle could account for
the success of science \cite{putnam7}. Note that although the putative
referents exist independently of the subject, their relationship to a given
theoretical term is not ascertained. Firstly, a theoretical term doesn't
necessarily refer to a real object (which will be called hereafter a
''physical entity''), but may refer to a {\em property} of an object, to
collective properties of an assembly of objects through model-reference, or
may not refer at all (examples will be given below). Most importantly, even
referring terms are approximate and revisable representations of physical
reality. Thus, although in successive theories theoretical terms do change,
they may still refer to the same physical entity, provided the description
given by the earlier theory is reasonably modified so that the entity
conserves its basic roles and properties: a new, more accurate theory brings
in novel features, but overlaps with the previous theory on a large domain
over which the older theory was known to be accurate. The typical example is
newtonian mechanics, which is still employed in many applications though it is known to be valid only in the non-relativistic limit. This does not mean
however that the older theory is necessarily a limiting case of the novel
one; for example Franklin concludes, in his historical study \cite{franklin}%
, that neutrinos retain to this day much of the original properties they had when
Pauli originally suggested the existence of such particles, although their
currently known properties (they come in three different varieties, and have
helicities) do not make them collapse to Pauli's neutrinos.

A few comments are in order. Our assumptions do not entail a
correspondence-theory of truth, by which the theoretical terms would be
endowed with a direct mapping unto physical reality. As a matter of fact,
the ontological perspective arising from a theory depends on whether (and
which) theoretical terms are given ontological reference. This last point
has always been discussed within the realist perspective and is intimately
related to what should be a realist account of truth\footnote{%
These discussions have over the years grown to incorporate sophisticated
technical arguments (see \cite{comp} for an introduction).}. We also see
that a realistic perspective doesn't necessarily call for a 'pictorial'
representation given in terms of intuitive or familiar categories known to
be useful in everyday life; in other words, there is no warrant that an
underlying ontology to a successful theory can be spelled out in these
categories. And it wouldn't be surprising for such pictorial
representations to fail for ontologies concerned with scales far away from the
biological realm where our common concepts have originated.

We finally wish to make some observations on Einstein's figure as
personifying realism. Einstein's mature position turns out to be at odds
with the previous remarks. Probably a remnant of his early Machian heritage,
Einstein starts constructing physics from the primary sensory perceptions we
experience, but insists into going beyond this phenomenological perspective:
physics is defined as a '{\em logical system of thought}' endowed with a
uniform logical basis \cite{einst}. He is not so much concerned about
whether the theoretical terms refer, but rather about whether a theory is
empirically adequate (which according to him is the main justification of the theory). As
regarding what type of elements an acceptable theory must contain, Einstein
does require the validity of a certain number of concepts from everyday
experience (causality, space-time representation), because for him, the
miracle is that the world is comprehensible, and his main concern is for
theories to give a plausible and intuitively understandable account
compatible with the available experimental data. This position was labeled
by Fine ''motivational realism'' \cite{fine86}, because its role is to give
a sense to Einstein's commitment to science, whereas stricto sensu, the
epistemological positions are closer to constructive empiricism than to
realism.

\section{Physical entities and invariance}

There undoubtedly are, for most physicists, real entities. Those are
detected and manipulated every day in research or development laboratories.
The properties of these physical entities are given within a physical theory
-- but do not entirely depend on the theory\footnote{%
The charge, mass or magnetic moment of the electron doesn't change when
considered within the Bohr-Sommerfeld theory or within the wave mechanics
arising from the Schr\"{o}dinger equation. The Dirac equation further adds a
new property (spin) to the electron, which previously needed to be
incorporated in an ad-hoc manner in the non-relativistic treatment. Other
properties which were subsequently discovered, such as the electron's role
in electroweak interactions, are taken into account by new theories which
encompass both the old and the novel properties.}. Traditionally, if not
intuitively, an entity has a certain number of depictable properties and its
evolution can be followed in space and time. Some of these properties are
contextual, other are invariants. In classical (including relativistic)
mechanics the contextual (usually frame-dependent) properties are related by
frame transformations; the invariants have the same value, independently of
the transformations. A property may then unambiguously be ascribed to an
entity, relative to a given reference frame, or equivalently, the invariants
under the given group of transformation may be constructed. This is why Max
Born insisted that '{\em the idea of invariant is the clue to a rational
concept of reality}' \cite{born53}.

This process is blurred in quantum mechanics. Quantum mechanically, an
object cannot generally be endowed with the properties it is supposed to
carry (this happens in compound systems, or for identical particles) -- a
situation known as the problem of objectification.\ And only the
time-evolution of the wave function of the physical system in configuration
space can be followed. Moreover, the wave function determines at most mean
values in conjunction with an observable, and probability amplitudes when
further associated with a definite quantized value (eigenvalue) of the
observable. Under general assumptions (without however introducing
additional hidden variables) it is not possible to attribute an individual
{\em eigenstate} of an observable to the system prior to a measurement
(strong objectification) nor even an {\em eigenvalue} of an observable to
the system (weak objectification) \cite{mittelstaedt} because observables do
not commute and wave-functions interfere (or put differently, because of the
wave character of the particles): there is no relation between the result of
a measurement and the one that would have been obtained from measuring
another observable in a different experimental setup, these measurements
being mutually exclusive and the result of a measurement not being known
with certainty. Hence this additional dependence on the experimental setup
(that is, on the projection basis in abstract spaces) which brings in a
novel type of hyper-contextuality unknown in classical mechanics.

However, the wave-function is invariant relative to the decomposition in
abstract Hilbert space, ie a unitary frame transformation between two
projection basis corresponding to two different observables does exist. This
'invariance' has been taken as an argument for the objectivity of the
wave-function itself -- the outcome of a measurement then appears as a
contextual property of an 'objective' physical system \cite{paty99}.
Notwithstanding, it must be recognized that invariance does not entail
reference. Furthermore, realism calls for more than a symbolic
transposition or representation: hence the problem of knowing what the wave function refers to
(see Sec. 4). Note that quite independently from the wave function's precise
role, invariant quantities are used in quantum mechanics to define the
natural kinds, that is to identify the physical entities. An electron is
characterized by its mass, charge, spin, whatever the contextual measurement
(as a matter of fact, these properties are indirectly measured from
different experimental setups). The elementary particles are distinguished
by a set of invariant properties. This is also true of compound systems as a
whole such as an atom, although further variables associated with the degree
of excitation appear. It is thus clear that these invariants are properties
of the physical entities that are manipulated experimentally.

\section{Theoretical terms and reference}

\subsection{Classical mechanics}

The problem of the reference of theoretical terms seems at first sight
straightforward in classical mechanics: either the objects exist, or they
don't. A simple well-known example is given by the debate on the existence
of the putative atoms postulated by the kinetic theory of gases towards the
end of the 19th century: were these atoms mere theoretical tools which saved
the phenomena, as the instrumentalists argued, or were they real but
at-the-time unobserved particles? A less simpler case concerns the existence
of the ether, which was postulated to exist because it was needed by the
ontological picture, although there wasn't any compelling theoretical term
which referred to it. A much more involved example can be found in the
criticism of the newtonian concept of force, which was much criticized from
its inception onwards as being obscure, while reference to less obscure
'active principles' was looked for. These examples illustrate the subtle
interplay between a physical theory, here classical mechanics, and the
ontological basis that goes with it, in the construction of an
interpretation.

In classical mechanics, the space-time (newtonian) formulation gives rise to
a simple reference procedure: the fundamental theory gives the properties of
material points $i$ as evolution of the position ${\bf x}_{i}(t)$ and
velocity ${\bf v}_{i}(t)$ of those points interacting through a function $V(%
{\bf x}_{1},{\bf x}_{2},...,{\bf v}_{1},{\bf v}_{2},...)$. The time
evolution is given by a differential equation of the type
\begin{equation}
m_{i}\frac{d^{2}{\bf x}_{i}}{dt^{2}}=-{\bf \nabla }_{i}V+\frac{d}{dt}\frac{%
\partial V}{\partial {\bf v}_{i}}
\end{equation}
which is readily amenable to an interpretation in terms of the action of a
force on the dynamics of the material points. Larger or complex systems are
investigated within models based on the fundamental theory, involving the
introduction of approximations; we speak of {\em model-reference}. The
relevant theoretical terms then refer to properties of physical entities
(usually some collective properties) through the model, which by definition
is itself an approximation. In statistical theories however, the terms do
not refer to properties of the physical entities: a statistical distribution
gives informations related to an ensemble of systems. It is an {\em epistemic%
} tool which at best refers to collective properties of a finite numbers of
identical systems, not to a property of an individual system.

Now, the most powerful and elegant formulation of classical mechanics based
on the theory of canonical transformations employs theoretical terms which
have a much more abstract and complicated relation to the physical entities.
An original set of independent coordinates and momenta $(q_{i},p_{i})$ is
transformed to a new set $(Q_{i},P_{i})$ by means of a generating function
\cite{goldstein80}.\ When all the $Q$'s and $P$'s are chosen to be constants
of motion, the generating function is the classical action $S(q_{i},P_{i},t)$
obtained by solving the Hamilton-Jacobi equation
\begin{equation}
H(q_{i},\frac{\partial S}{\partial q_{i}},t)+\frac{\partial S}{\partial t}=0,
\end{equation}
where $H$ is the Hamiltonian expressed in terms of the independent
coordinates, and the original momenta are retrieved by using
\begin{equation}
p_{i}=\frac{\partial S}{\partial q_{i}}.
\end{equation}
Of course other generating functions expressed in terms of different sets of
independent coordinates may be chosen (eg the function labeled $%
F_{4}(p_{i},P_{i},t)$ by Goldstein \cite{goldstein80}; in this
representation, the $p_{i}$'s are the independent coordinates, and the
positions are retrieved by the relation $q_{i}=-\partial _{p_{i}}F_{4}$).
Theoretical emphasis is put on canonically invariant quantities, such as the
Poisson bracket of two quantities $a(q,p)$ and $b(q,p)$, defined by
\begin{equation}
\left\{ a,b\right\} \equiv \sum_{i}\frac{\partial a}{\partial q_{i}}\frac{%
\partial b}{\partial p_{i}}-\frac{\partial a}{\partial p_{i}}\frac{\partial b%
}{\partial q_{i}}.
\end{equation}
The effect of a transformation can then be written in terms of Poisson
brackets; for example the time evolution of $a(p,q)$ is given by
\begin{equation}
\frac{da}{dt}=\left\{ a,H\right\} +\frac{\partial a}{\partial t},  \label{5}
\end{equation}
where $H$ is now expressed in terms of $p$ and $q$.

What does the Poisson bracket or the classical action refer to?
There is clearly no place for them in the ontology underlying
classical mechanics; these terms are not properties of physical
entities. Relative to the ontology of classical mechanics, they
are epistemic terms from which we can extract the dynamics of the
system (the action -- a non-local quantity -- gives all the
possible trajectories compatible with the mechanical system, the
bracket gives the evolution generated by an infinitesimal
canonical transformation). The propagation of the action in
configuration space can be studied for its own sake (for example
the surface of constant $S$ forms a wave-front normal to the
trajectories propagating with velocity $E/p$).\ But this doesn't
make the action a referring term, although we know how to
associate this term with the trajectories of physical entities.

\subsection{Quantum mechanics}

The main problem in understanding quantum mechanics is centered on
the meaning of the wave function. Ontological interpretations
hinge on the task of referring the wave-function to some
recognizable physical phenomenon, and there is clearly no
agreement on this point. Schr\"{o}dinger's original position,
abandoned soon afterwards, was to envision the wave function as
existing in real space.\ The well-known 'many worlds
interpretation' is a consistent manner of taking the wave function
at face-value. A more elaborate form of combining projective
decompositions of the wave-function evolution is achieved by the
consistent histories interpretation; however, the ontological
status of the histories (each one of them is consistent, but two
histories are generally incompatible accounts of the same
phenomena) is open to question. In the de Broglie-Bohm account of
quantum mechanics, the wave function is referred to a real
physical field existing in configuration space; this space
therefore acquires an ontological existence. More recently, the
introduction of non-destructive measurements has lead certain
authors \cite{aharonov} to ascribe to the wave function the role
of referring to a time-averaged property of a particle.

To tackle the problem of the reference of the wave function, let us start by recalling that there are three
well-established connections between quantum and classical theoretical
terms\footnote{%
We leave aside the phase-space formulation of quantum mechanics of
the Weyl-Wigner-Moyal type, where the connection between quantum
and classical terms is more complex.}. First, there is the formal
correspondence between a classical Poisson bracket and the
quantum-mechanical commutator, a procedure known as canonical
quantization. For example the classical relation $\left\{
q_{i},p_{j}\right\} =\delta _{ij}$ has the quantum counterpart
($i\hbar )^{-1}\left[ \hat{q}_{i},\hat{p}_{j}\right] =\delta
_{ij},$ where the hat denotes an operator, and Eq. (\ref{5}) has
the counterpart
\begin{equation}
\frac{d\hat{a}}{dt}=\frac{1}{i\hbar }\left[ \hat{a},\hat{H}\right] +\frac{%
\partial \hat{a}}{\partial t},
\end{equation}
that is the evolution of a quantum-mechanical operator $\hat{a}(t)$ in the
Heisenberg picture. Second, the path-integral approach gives the transition
amplitude of the evolution operator in the time-interval $t_{f}-t_{0}$ as
\begin{equation}
\left\langle q_{f},t_{f}\right. \left| q_{0},t_{0}\right\rangle =\int {\cal D%
}(q)\exp \left[ i\int_{t_{0}}^{t_{f}}dtL(q,\dot{q})/\hbar \right]
\label{10}
\end{equation}
where the integral of the Lagrangian $L$ is of course $%
S(q_{f},t_{f};q_{0},t_{0})$, the classical action. Finally, the so-called
WKB theory gives the semiclassical approximation of the wave function as
\begin{equation}
\psi (q_{f},t_{f})=A(q_{0},t_{0})\left| \det \frac{\partial q_{0}}{\partial
q_{f}}\right| ^{1/2}\exp \left[ iS(q_{f},t_{f};q_{0},t_{0})/\hbar -\mu \pi
/2\right] ;  \label{11}
\end{equation}
$A^{2}$ represents the classical probability density and $\mu $ is the
Maslov index keeping track of the caustics (the points in phase-space where
the semiclassical approximation is not valid) along the trajectory.

Note that in order to obtain Eqs. (\ref{10}) and (\ref{11}) assumptions
which are seldom mentioned and which are not compelling within quantum
mechanics must be made. In Eq. (\ref{10}), although the different variables $%
q_{k}$ that appear in the measure ${\cal D}(q)$ are {\em independent}
variables of integration, the classical action is obtained only by
forcefully identifying $q_{k+1}-q_{k}$ with $\dot{q}_{k}dt$ \cite{swanson94}%
. In an analog manner $S$ is only one of the infinite number of phases
yielding the semiclassical wave function (\ref{11}) (quantum mechanics only
imposes that all these phases obey a nonlinear equation \cite{matzkin2001},
provided the amplitude $A$ changes accordingly). These stronger assumptions
must be made if any connection between quantum mechanical and classical are
to be made at all.\ These assumptions are known to be consistent; for
example in many systems for which $\hbar $ can be scaled, preferential
propagation of the wave function along classical periodic trajectories has
been observed as $\hbar \rightarrow 0,$ a finding which has sparked progress
in many branches of semiclassical physics \cite{brack badhuri97}.

What is striking is that the quantum-mechanical relations are made-up with
quantities that classically have an {\em epistemic non-referring}
interpretation\footnote{%
There is of course no warrant that the same interpretation needs to hold
within quantum mechanics. In fact, there are historical examples and
counterexamples: Ptolemy's epicycles and cristalline spheres never came to
refer to anything, whereas the energy, which was regarded as an abstract
mathematical tool, became a central referring property by the end of the
19th century.}. The operators alone are not sufficient to obtain the
dynamics; they must be associated with a wave function to obtain eigenvalues
or mean values. The wave function is generally obtained from the
eigenfunctions of the Hamiltonian, and in any case depends on the
interactions between the particles: for example the wave function of, say an
electron in a magnetic field is quite different from the wave function of an
electron in a Coulomb field, although in both cases the wave function is
related to the same physical entity that has the same invariant properties.
The fact that the wave function conveys different dynamical information
while supposedly referring to the same object is again in favour of an
epistemic interpretation. The wave-function invariance (in Hilbert space)
argument mentioned in Sec. 3, according to which the objectivity of the wave
function is increased by this invariance, was seen to have its counterpart
in classical mechanics, where different representations for the abstract and
non-referring generating functions can be chosen. Other arguments which
render difficult the task of referring the wave function include the
unchallenged probability postulate and thus the need for a normalization
factor, the irrelevance of a global phase in the wave function and the
non-invariance under Galilean transformations of the local wavelength of the
wave function. The most problematic point, however, remains the reduction of
the wave function during a measurement process. It seems unlikely that any
decisive progress on the meaning of the wave function will be made before a
satisfying solution to the measurement problem will be found.

\section{Implications}

We have argued that within a realist perspective, it is not compelling to
refer the wave function to a physical entity, since not all theoretical
terms refer to physical entities or their properties.\ We examine the
implications on the interpretation and the teaching of quantum mechanics.

\subsection{Interpretation}

Whereas empiricists and instrumentalists consider quantum mechanics as a
masterpiece to validate their arguments, realist interpretations have been
obscured by two points. First,{\em \ }a certain number of preconceived
epistemological constraints that were imposed as basic requirements in any
realistic account (an example is Einstein's position, as sketched in Sec.\
2); for example it is often thought that realism imposes an isomorphism
between the theoretical terms and nature (eg, objectivity, understood in the
precise sense of relating a theoretical term with an element of reality with
unit probability -- as it appeared in the EPR paper \cite{EPR} -- is
believed to be a basic prerequisite for any theory compatible with realism).
The upshot is that these epistemological constraints aim at bypassing any
epistemic barrier between our theories and reality, thus implicitly assuming
a pre-structured reality and a correspondance theory of truth.

The second point concerns a preconceived ontology; a realist account is
often thought to be necessarily spelled out in apparently familiar terms,
although there is a price to pay because the ontology must be modified to
account for the novel phenomena. A demonstrative example is the de
Broglie-Bohm approach to quantum mechanics \cite{holland}: a wave -- a real
physical field propagating in configuration space -- guides the particles
along in principle unobservable trajectories. Originally devised to restore
a continuity between classical descriptions and quantum phenomena -- it was
later claimed that it would avoid '{\it arbitrary dichotomies ... between
evident (macroscopic) realism and quantum (microscopic) nonrealism}' \cite
{bohm hiley85} --, Bohmian mechanics has been led to ascribe to the putative
entities a haul of counterintuitive properties: particles are detected even
though no trajectories go nearby the detector, properties such as the mass
of the particles are delocalized, configuration space has sometimes been
claimed to be more fundamental than our 3+1 D space-time, classical
trajectories are seldom obtained in the classical limit. Such
peculiarities are given ad-hoc assumptions if necessary so as to yield the
standard or experimental results. In view of this situation, it might appear
as a paradox for Bohmian mechanics to be coined {\em the} interpretation of
quantum-mechanics compatible with realism, even sometimes as the main
counter-argument used by many realist philosophers with an interest in the
interpretation of quantum mechanics \cite{norris2000,bub97}\footnote{%
This forms part of a more general tendency some philosophers have
which is to use any interpretation giving the wave-function an
ontological existence as an argument for realism.}.

The consistent histories approach \cite{omnes92} offers a quite
different perspective regarding the interplay between
epistemological and ontological constraints. From our point of
view, the main feature comes from the existence of alternative and
mutually exclusive consistent histories. For example in a
beam-splitter (this example is discussed in \cite{griffiths98}),
the neutron, at some intermediate time $t_{1}$ before being
detected, is in either of the two channels if it is described by
two different histories of a certain family; a history belonging
to another family states that at $t_{1}$ the wave-function is in a
superposition state. What can be the ontological value of this
"many-picture" formalism? Each history taken individually makes
sense from an epistemic standpoint (and the history thereby allows
us to intuit the physical evolution of the system in-between
observations), but \emph{any} of the alternative and complimentary
pictures may be taken as valid. The properties of the physical
phenomena is thus context-dependent through the history that is
chosen for the description, and in this respect, the consistent
histories interpretation does not shed new light on the properties
and objectification problems briefly mentioned in Sec. 3. Maybe
our brains lack the categories in which reference could be
expressed (we would then be in presence of an internal biological
epistemic barrier), and the problem of objectification can
therefore be dismissed: there are real physical entities, but as
far as we can know, their properties can only be described
relative to a given history (compatible with a given experimental
setup). However, we noted there are invariant, context-independent
properties, and we suggested that the wave-function was "made-up"
of classically non-referring terms. Moreover an actual measurement
performed at an intermediate time yields a unique result, and we
know that a unique classical macroscopic description arises from
the diverse quantum histories. This is why, in our view, the task
of constructing an ontological picture from the physical theory
hinges on the resolution of the measurement problem -- the
quantum-mechanical description of
classical phenomena\footnote{%
There have been claims (eg \cite{omnes92}), that at least for
practical purposes, the existence of classical properties is well
understood, but there is no agreement even within the consistent
histories perspective \cite{kent00}.}.

We have seen that realism doesn't entail a correspondence theory
of truth by which every element of reality could be mapped onto
theoretical terms. Moreover, realism is not concerned by adapting
a physical theory to a preconceived ontological basis: as
celebrated examples in the history and philosophy of science have
taught us, this can only be done by developing a plethora of
ad-hoc hypothesis, increasingly non-referring and unphysical. If,
as we have argued, realism is based on referring theoretical terms
to physical entities, there is no place for in principle
unobservable properties, because observation is the warrant of
referentiability, ie it is only by repeating and combining
observations that the referents can be accessed \cite{agazzi00}.
Although we know how to recognize real physical entities, through
their invariant properties, it is not clear what the furniture of
the quantum world is.

\subsection{Teaching}

If, following Feynman, {\em nobody} does understand quantum theory (for a
more nuanced review of the present situation, see \cite{laloe2001}),
particular difficulties in the teaching of quantum physics are to be
expected. Indeed, recent investigations in the students' understanding of
quantum phenomena have not surprisingly diagnosed profound conceptual
difficulties \cite{mash99,ireson99,vokos etal00}. More surprisingly however,
opposite conclusions have been drawn from these studies: in one case \cite
{ireson99} it is recommended to avoid any reference to classical physics and
to dual quantum/classical descriptions, such as the Bohr atom, in favour of
statistical interpretations of observed phenomena. In another case \cite
{mash99}, it was found that the difficulties arise from the ontological and
epistemological status that students ascribe to the theoretical terms, and
it was suggested to develop 'mental models' that would reconcile quantum and
classical physics. More generally the findings indicate conceptual
difficulties that later persist when more advanced material is studied; only
specific tutorials based on constructing concepts and '{\em relating the
formalism of physics to the real-world phenomena}' \cite{vokos etal00} were
found to be efficient.

If conceptual problems deserve conceptual treatments, it then appears that
the specific problems raised by quantum mechanics can hardly be understood
without going into a more general inquiry on the relationship between
theories and reality. We have argued that an approach from the realist
standpoint of reference is well-suited in order to understand this
relationship. Globally, this suggests that introductory courses to the
philosophy of science would help the students in their confrontations with
the inevitable conceptual problems that arise when trying to understand what
is quantum mechanics and what is quantum reality \cite{phil}.

\section{Conclusion}

Realism leads us to believe that physical theories refer to real
physical entities. This doesn't mean that every theoretical term
must refer. We have given examples of theoretical terms that have
an epistemic non-referring signification in classical mechanics.
We have further seen that quantum-mechanical theoretical terms can
be constructed with these classical epistemic terms. The tension
between the invariant properties of the physical entities and the
contextual nature of the wave function suggests that a realist
interpretation will come through with the solution to the
measurement problem, rather than by imposing preconceived
epistemological or ontological constraints.

\ack Financial support from the European Commission's IHP-MCIF
Programme is acknowledged.

\end{document}